# A new twisting phase transition in layered materials

Ze Liu[1]

[1] Department of Engineering Mechanics, School of Civil Engineering, Wuhan University, Wuhan, Hubei 430072, China.

Correspondence and requests for materials should be addressed to Z.L. (email: ze.liu@whu.edu.cn).

**Twisted two-dimensional (2D) layered materials exhibit many novel and unique phenomena, such as insulation and superconductivity transition, and superlubricity. However, the effect of twisting on these phenomena remains unclear. A key challenge is the lack of linkage between the twisting and the material properties. Here, twisting induced moiré pattern is directly correlated with the configuration entropy of the bilayer graphene, based on which a new twisting phase transition in layered materials is uncovered. Most critically, the twisting phase transition is found size dependent. The diversity of phase transition among the AB stacking and the twisted phases could provide insights into the twistronics and twistribology.**

Two-dimensional (2D) materials provide a platform to explore novel electronic [1,2], optical [3] and tribology [4-9] properties, and to create new materials by constructing heterostructures [10,11] with a variety of building units, ranging from semimetallic graphene [12], insulating hexagonal boron nitride [13] to semiconducting black phosphorus [14-16]. Beyond the construction of heterostructures, it was predicted that the electronic structure of a bilayer graphene would significantly change by simply twisting the bilayer graphene away from the AB stacking [17,18]. Recently, it was indeed experimentally observed that both the superconducting and the insulating states can appear in bilayer graphene when twists the bilayer graphene to an angle of ~1.1° (called "magic angle") [19,20]. Such a phenomenon is attributed to the strong correlation between electrons in the "magic angle" graphene [21,22]. However, the physical mechanism behind these observations has remained unknown [23]. And it is not clear how does the geometric moiré pattern correlate with the measurable physical quantities.

Here, we report the correlation of geometric moiré pattern with the configuration entropy of twisted bilayer graphene. By introducing this configuration entropy into the Gibbs free energy, we show the existing of diversity of phase transition in bilayer graphene. For convenience, we classify the twisted bilayer graphene into two types of phases, i.e. superlattice and disorder phases, where in the former, the positions of atoms in one moiré unit-cell are strictly repeated in other moiré unit-cells, and in the latter, they are not. We found that there exists a large number of transition among the phases of AB stacking, superlattices, and disorders. Most critically, we found that such transitions are significantly dependence on the sample size.

Generally, moiré patterns form when two periodic patterns are overlaid with a relative twisting angle [17]. For a bilayer graphene, once relatively rotating the contacted two graphene layers with a twisting angle of $\theta$, the size of moiré unit-cells ($a_m$) is [18]

$$a_m = \begin{cases} \sqrt{3}a & , \theta = 0 \\ \frac{\sqrt{3}a}{2\sin(\theta/2)} & , 0 < \theta \leq 30° \end{cases} \quad (1)$$

where $a$ is the C-C bond length of graphene lattice. Because $a_m(\theta)$ possesses the rotational symmetry of 30º and 60º. Thus, we only need to consider $a_m(\theta)$ in the range of $0 \leq \theta \leq 30°$, We argue that such a purely geometric moiré pattern actually reflects the disorder degree in the twisted bilayer graphene. For simplicity, we consider one superlattice phase with twisting angle of $\theta = 13.173°$ (Fig. 1a-d) for example. It is needed to point out that the number of such superlattices is infinite [18]. For any given twisting angle, the size of the moiré unit-cell (the orange lines enclosed region in Fig. 1a-d) is determined by eq. 1 and the number of atoms in one moiré unit-cell (within one graphene layer) is thus $n = \rho A_\theta$, where

$\rho = \frac{4}{3\sqrt{3}a^2}$ is the in-plane density of carbon atoms. $A_\theta = \frac{\sqrt{3}}{2}a_m^2$ is the area of one moiré unit-cell. It is noted that there are two types of atoms in one moiré unit-cell, such as A, B… and C, D…(Fig. 1a-d). If we translationally move the upper graphene among the positions of atoms of the same type, such as from point A to B (Fig. 1b) or from point C to D, it is obvious that the system internal energy doesn't change because the atomic moiré patterns are the same, where the atomic moiré pattern is defined to distinguish the geometric moiré pattern, and it means the positions of atoms in the moiré unit-cells are strictly repeatable. Furthermore, within any of the moiré unit-cells, the potential energies of the carbon atoms (of the same type, $g(r)$) with the other graphene are different since their position vectors (*r*) are different (Fig. 1e-f), where *r* is the position vector of an atom relative with the central point of a carbon six-membered ring in the bottom graphene. It is obvious that $\varepsilon_{AB} \leq g(r) \leq \varepsilon_{AA}$, where $\varepsilon_{AB}$ and $\varepsilon_{AA}$ denote that the atom is in AB stacking and AA stacking with the bottom graphene (Fig. 1e-f), respectively. Therefore, we can label them as *A*, *B*, *C*… (Fig. 1a-d) and distinguish them. For the cases of superlattices, the positions of carbon atoms in one moiré unit-cell are exactly repeated in all other moiré unit-cells (such as *A'*, *B'*, *C'* in Fig. 1a-d). The disorder degree associated with the twisted bilayer graphene is thus determined by how many different configurations (or micro-status) can the twisted bilayer graphene have, under the conditions of the same internal energy and the same twisting angle. To keep the twisting angle, the configurations must possess translationally symmetry. We can herein translationally move the top graphene among the points of *A*, *B*, *C*… For example, we can translationally move the top graphene from point A to B (Fig. 1b), the configuration is obviously different (the relative positions of the labeled atoms in the geometric moiré patterns are different), but the internal energy of the system is the same. However, if we translationally move the top graphene from position A to C (Fig. 1c), the system energy changes because the atomic moiré pattern changes. In addition, if we translationally move the top graphene from position A to A' (Fig. 1d), it is obvious that it is not a new configuration (the same as Fig. 1a). Therefore, for any atom in one moiré unit-cell, which represents the top graphene (rigid approximation), there are $n/2$ different energy levels by the translation operation. The number of micro-status of a bilayer graphene is thus

$$\Omega = n/2 \tag{2}$$

And the corresponding entropy associated with the twisting is

$$S = k \ln \Omega = k \ln(n/2) \tag{3}$$

where $k = 1.38 \times 10^{-23}$ J/K is the Boltzmann constant. Especially, for the AB stacking bilayer graphene, its configuration entropy is $S = k \ln 2$, which means that the entropy of pure and perfect bilayer graphene is not zero at the absolute temperature of zero.

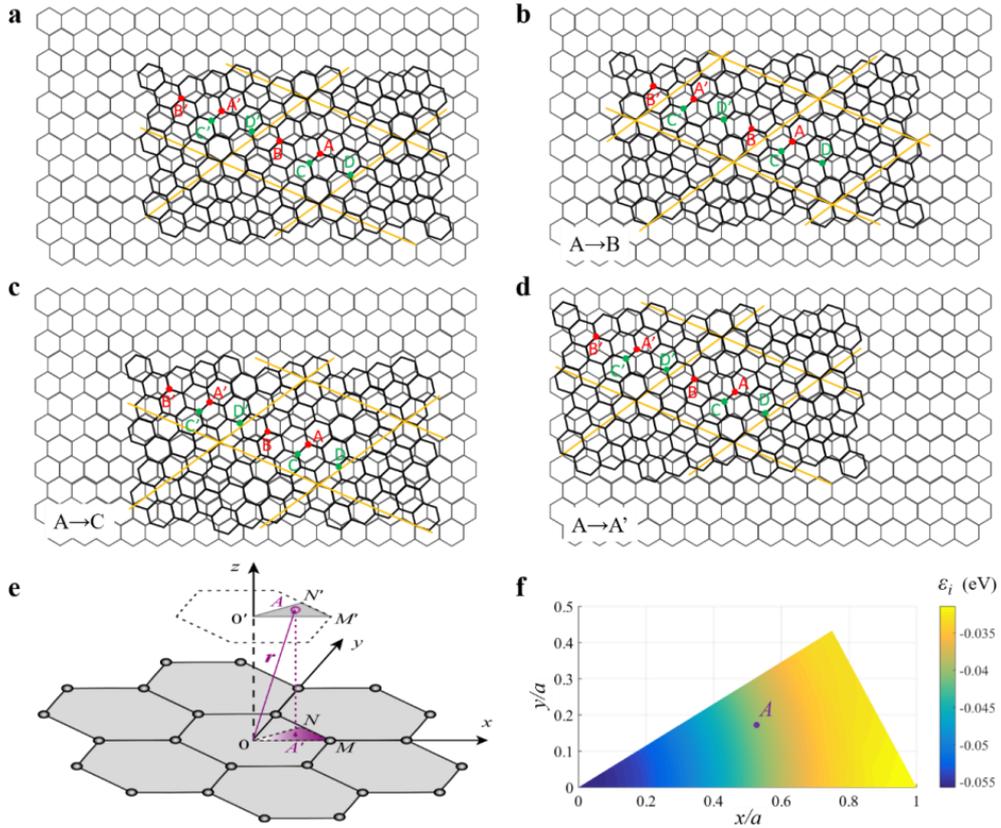

Fig. 1. Micro-status in twisted bilayer graphene. (a)-(d) A graphene flake with finite size twists on a large graphene, where the twisting angle is $\theta = 13.173°$, corresponding to a superlattice phase (the positions of atoms in the moiré unit-cells (orange lines enclosed regions) are exactly repeatable). (b) Translationally move the top graphene from point A (a)

to B. (c) Translationally move the top graphene from point A (a) to C. (d) Translationally move the top graphene from point A (a) to A'. (e) The potential energy of one carbon atom moves on a graphene with constant normal distance depends on its position vector $r$. The energy level of the atom $A$ is bounded by the AB stacking ($\varepsilon_{AB}$, $A$ located at 0') and the AA stacking ($\varepsilon_{AA}$, $A$ located at M') and is calculated by using the (Kolmogorov–Crespi (KC) potential [24] (f).

As for the disorder phases where the positions of atoms in one moiré unit-cell are not exactly repeated in other moiré unit-cells, eqs. (2) and (3) still approximately hold true. This is because the energy distribution of atoms, $g(r)$, in the moiré unit-cells is the same (Fig. 1f) no matter whether the positions of atoms in the moiré unit-cells have translational symmetry or not. Therefore, for a given energy range $(\varepsilon_i, \varepsilon_i + d\varepsilon_i]$, where $\varepsilon_i \in [\varepsilon_{AB}, \varepsilon_{AA}]$, there are the same number of atoms in different moiré unit-cells with this energy level. Then if the atoms in one moiré unit-cell traversed all the energy levels in $g(r)$, the atoms with the same energy level in other moiré unit-cells also traversed all the energy levels in $g(r)$. In other words, the micro-status of the twisted bilayer graphene is determined by the number of atoms with different energy levels in one moiré unit-cell.

After correlating the moiré pattern induced by twisting with the configuration entropy of a bilayer graphene, we can thus link the twisting with the material properties of the bilayer graphene. In general, all materials exhibit various forms that are characterized by different physical properties such as density or molecular structure. Virtually, all phase transitions occur under constant pressure and temperature, can be best described by the Gibbs free energy. Two phases in equilibrium under such conditions possess equal Gibbs free energy. The Gibbs free energy ($G$) of a bilayer graphene is

$$G = U + PV - TS \qquad (4)$$

where $U$, $P$, $V$ and $T$ are the internal energy, the pressure, the volume and the temperature of the system, respectively. For bilayer graphene without external forces, $P$ and $V$ are constant and we can remove them from eq. (4) without affecting the comparison of the Gibbs free energy of different phases. By substituting eq. (3) and $n = \rho A_\theta$ into eq. (4), we finally have

$$G = U(\theta) + kT \ln\left(4\sin^2\left(\frac{\theta}{2}\right)\right) \qquad (5)$$

The twisting angle dependent internal energy in eq. (5) can be easily obtained through the molecular simulations by using the KC potential [24,25]. Typical results for one graphene with radius of $R = 1$ nm twisting on a large graphene are shown in Fig. 2a, where the initial stacking is the AB stacking, the red dots represent the superlattice phases, and the other twisting angles correspond to the disorder phases. The superlattice phase associated twisting angle is given as[18] $\cos\theta_i = (3i^2 + 3i + 0.5)/(3i^2 + 3i + 1), i = 0, 1, 2 ...$ Based on eq. (5), the stability of these phases (the AB stacking, the superlattice, and the disorder phases) can be quantified by comparing their Gibbs free energy. Typical phase transition between the AB stacking and the superlattice S1 (with twisting angle of $\theta \approx 2.005°$) is shown in Fig. 2b, which clearly tells that the critical transition temperature is $T_c \approx 178$ K, below which the AB stacking is the stable phase while the superlattice S1 becomes more stable when $T > T_c$. It is needed to point out that this conclusion is on the basis of equilibrium phase transition, if dynamic process is considered, the superlattice S1 could transform to the AB stacking even under low temperature due to the interlayer frictionless sliding (superlubricity) [4,7,8,26].

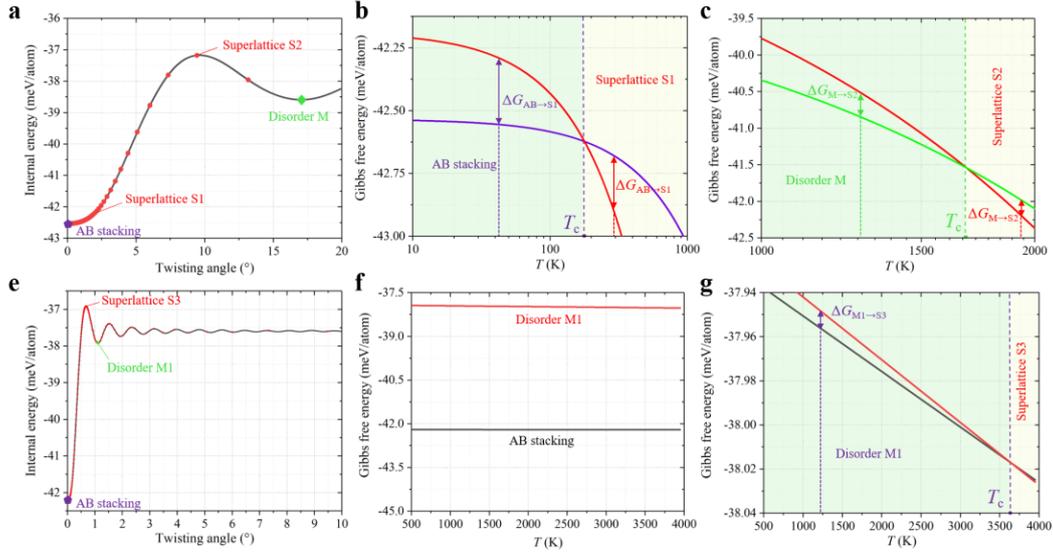

Fig. 2. Phase transition among the AB stacking and the twisted phases. (a) The calculated internal energy of a bilayer graphene, where a graphene flake with radius of $R = 1$ nm twists on a large graphene. (b) Comparison of the Gibbs free energy between the phases of the AB stacking and the superlattice S1 in (a). (c) Comparison of the Gibbs free energy between the phases of disorder M and the superlattice S2 in (a). (e)-(f) Typical phase transition in a bilayer graphene with radius of $R = 15$ nm.

Besides the phase transition between the AB stacking and the superlattice phases, there could exist phase transition between the superlattice and the disorder phases. Typical phase transition between the phases of the superlattice S2 and the disorder M is shown in Fig. 2c, it is obvious that the disorder M (local internal energy minimum) is more stable when $T < T_c \approx 1680$ K, which indicates that though the AB stacking is the global internal energy minimum, high temperature annealing is required to make twisted bilayer graphene trapped in local internal energy minima transforming to the AB stacking. Interestingly, we observed that as the sample size increasing, the internal energy landscape becomes more flat, but more and more local minimum points appear (Fig. 2d). Similarly, we found that the twisted bilayer graphene trapped in the local internal energy minimum ($\theta \approx 1.11°$) can only be transformed to the global minimum (AB stacking) with the aid of high temperature annealing ($\geq 3640$

K) (Fig. 2e-f). These predictions agree well with experiments. For example, it was found that any heat annealing after fabrication of twisted bilayer graphene encapsulated with hexagonal boron nitride flakes tends to relax the twisted bilayer graphene to the AB stacking at high temperatures [19]. More experimental evidence is from the thermal induced graphene rotation on hexagonal boron nitride [27], where the authors found that the thermal-induced rotation phenomenon is universal at elevated temperatures by examining tens of samples and they didn't observe any rotation of these graphene flakes for at least one month at room temperature, which indicates that there must exist a critical transition temperature, but unfortunately it was not given in the experiments. Moreover, the authors observed that the larger graphene flakes can be stable at twisted angles more near 0°, which can also be explained by our theory: as the flake size increasing, more and more local internal energy minima appear in the internal energy landscape and the local minimum adjacent to the global minimum becomes closer and closer (Fig. 2a and d). Therefore, the twisted sample can be trapped in the local internal energy minima as experimentally observed, unless the annealing temperature is high enough.

Another striking feature of our theory is that it predicts that there is diversity of phase transition in the bilayer graphene system, which may provide some insights into the recently observed interlaced modes between the insulation and the superconductivity in "magic angle" graphene [28]. It was found that such interlaced modes are more complex than predicted [28]. In addition, the reported magic angle range also extends from the initial ~1.1° to 0.93°[29]. We speculate that the above phenomena may originate from the infinite possible phase transitions in the system and the difference in the "magic angle" could be from the observed size effect here. It must be noted that our analysis assumes that the deformation of the bilayer graphene is negligible and the internal energy of the system is only twisting angle dependent, while in experiments [19,20,28], the twisted bilayer graphene is generally constrained by substrates or electrodes. The variation of the twisting angle during phase transition can be from strain but it will be small. In addition, the internal energy of twisted bilayer graphene in experiments was also tuned by the exerted voltage, which is not considered here.

Besides the size effect in the twisting angle dependent internal energy landscape, the twisting induced configuration entropy is also size dependent (eq. (3)). On the one hand, for a given twisting angle (or $a_m$), when the size of a bilayer graphene is smaller than $a_m$, the configuration entropy is linearly increasing as the number of atoms in the graphene flake, or equivalently, linearly scales with the sample size. However, if the flake size continuous increase to be larger than $a_m$, its configuration entropy will be size independent. On the other hand, if the flake size is constant, taking the bilayer graphene with radius of $R = 15$ nm for example (Fig. 2d), when the twisting angle is smaller than ~0.50° (point A in Fig. 3a), $a_m$ is always larger than the sample size, so the configuration entropy of the bilayer graphene is constant (determined by the number of atoms in the sample, Fig. 3a). While for $\theta > 0.50°$, the configuration entropy continuous decreases since $a_m$ decreases as $\theta$. By comparing the Gibbs free energy between the AB stacking and the twisted phases, we get the general phase transition temperature as

$$T_c = \frac{U(\theta) - U(0)}{k \ln(n/4)} \quad (6)$$

Based on eq. (6), the critical temperature for the transition between the AB stacking and any twisted phases can be obtained. Typical results for the bilayer graphene with $R = 15$ nm is shown in Fig. 3b. It is clear that to induce phase transformation from the AB stacking to a twisted phase, the required temperature drastically increases as the twisting angle, which explains why the AB stacking is the most common phase in experiments.

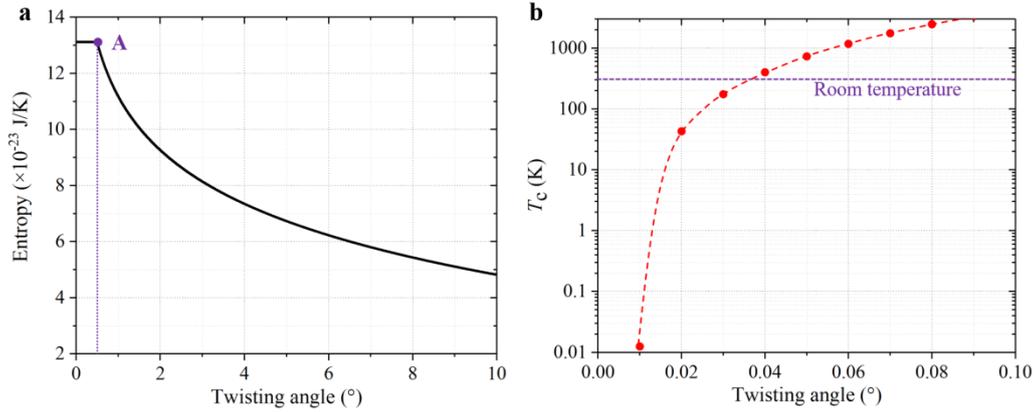

Fig. 3. Twisting angle dependent configuration entropy and the phase transition temperature, where a graphene with radius of $R = 15$ nm twists on a large graphene. (a) Configuration entropy of the bilayer graphene versus the twisting angle. (b) Calculated critical temperature for the phase transition between the AB stacking and the twisted phases.

To summary, by correlating the micro-status of a bilayer graphene with the geometric moiré patterns, we formulated the twisting induced configuration entropy in bilayer graphene. Introducing the configuration entropy into the Gibbs free energy allows us to uncover new phase transitions among the AB stacking and the twisted phases. We are interested to find that such phase transitions significantly depend on the sample size. In addition, our theory predicts that twisted bilayer graphene is not stable at high temperatures, which agrees well with experiments, where the authors found that the prepared twisted bilayer graphene and twisted graphene/h-BN are not stable and they will transform to the AB stacking at high temperatures [19,27]. Most critically, our theory reveals that there exists diversity of phase transitions in twisted bilayer graphene, which may provide insights into the recently observed insulation and superconductivity in "magic angle" graphene.

**Acknowledgements.** This research was supported by the funding from National Natural Science Foundation of China (11872284 and 11602175), and Wuhan Science and Technology Bureau of China (2019010701011390). Correspondence and requests for materials should be addressed to Dr. Ze Liu (ze.liu@whu.edu.cn)